\documentclass[12pt,prd,aps,amssymb,amsmath,tightenlines,showpacs,a4paper,nofootinbib]{revtex4}
\def\bea{\begin{eqnarray}}
\def\eea{\end{eqnarray}}
\def\beax{\begin{eqnarray*}}
\def\eeax{\end{eqnarray*}}
\def\half{\frac{1}{2}}
\def\quarter{\frac{1}{4}}
\def\vf{\varphi}
\usepackage{graphics}
\begin{document}
\title{Perturbation Theory for $PT$-Symmetric Sinusoidal Optical Lattices at the Symmetry-Breaking Threshold}

\author{H.~F.~Jones\email{h.f.jones@imperial.ac.uk}}

\affiliation{
Physics Department, Imperial College, London SW7 2BZ, UK\\}
\date{\today}

\begin{abstract}
The $PT$ symmetric potential $V_0[\cos(2\pi x/a)+i\lambda\sin(2\pi x/a)]$ has a completely real spectrum for $\lambda\le 1$, and begins to develop complex eigenvalues for $\lambda>1$. At the symmetry-breaking threshold $\lambda=1$ some of the eigenvectors become degenerate, giving rise to a Jordan-block structure for each degenerate eigenvector. In general this is expected to give rise to a secular growth in the amplitude of the wave. However, it has been shown in a recent paper by Longhi, by numerical simulation and by the use of perturbation theory, that for an initial wave packet this growth is suppressed, giving instead a constant maximum amplitude. We revisit this problem by developing the perturbation theory further. We verify that the results found by Longhi persist to second order, and with different input wave packets we are able to see the seeds in perturbation theory of the phenomenon of birefringence first discovered by Makris et al.

\end{abstract}

\maketitle

\section{Introduction}
The study of quantum mechanical Hamiltonians that are $PT$-symmetric but not Hermitian\cite{BB}-\cite{AMh} has recently found an unexpected application in classical optics\cite{op1}-\cite{op9}, due to the fact that in the paraxial approximation the equation of propagation of an electromagnetic wave in a medium is formally identical to the Schr\"odinger equation, but with different interpretations for the symbols appearing therein. It turns out that propagation through such a medium exhibits many new and interesting properties, such as power oscillations and birefringence.
The equation of propagation takes the form
\bea\label{opteq}
i\frac{\partial\psi}{\partial z}=-\left(\frac{\partial^2}{{\partial x}^2}+V(x)\right)\psi,
\eea
where $\psi(x,z)$ represents the envelope function of the amplitude of the electric field, $z$ is a scaled propagation distance, and $V(x)$ is the optical potential, proportional to the variation in the refractive index of the material through which the wave is passing. A complex $V$ corresponds to a complex refractive index, whose imaginary part represents either loss or gain. In principle the loss and gain regions can be carefully configured so that $V$ is $PT$ symmetric, that is $V^*(x)=V(-x)$. There is also a non-linear version of this equation, arising from sufficiently intense beams, where there is an additional term proportional to $|\psi|^2\psi$. However, for the purposes of this paper we shall limit ourselves to the linear case.

A model system exemplifying some of the novel features of beam propagation in $PT$-symmetric optical lattices uses
the sinusoidal potential
\beax\label{Vsin}
V=V_0\left[\cos(2\pi x/a)+i \lambda \sin(2\pi x/a)\right]
\eeax
This model has been studied numerically and theoretically in Refs.~\cite{op3,op6,op7}. The propagation in $z$ of the amplitude $\psi(x,z)$ is governed by the analogue Schr\"odinger equation\ref{opteq}, which for an eigenstate of $H$, with eigenvalue $\beta$ and $z$-dependence $\psi\propto e^{-i\beta z}$ reduces to the eigenvalue equation
\bea\label{H}
-\psi''-V_0\left[\cos(2\pi x/a) + i \lambda \sin(2\pi x/a)\right]\psi= \beta\psi\ .
\eea
 It turns out that these eigenvalues are real for $\lambda\le 1$, which corresponds to unbroken $PT$ symmetry, where the eigenfunctions respect the (anti-linear) symmetry of the Hamiltonian. Above $\lambda=1$ complex eigenvalues begin to appear, and indeed above $\lambda\approx 1.77687$ all the eigenvalues are complex\cite{Midya}. Clearly one would expect oscillatory behaviour of the amplitude below the threshold at $\lambda=1$ and exponential behaviour above the threshold, but the precise form of the evolution at $\lambda=1$ is less obvious. At first sight one would expect linear growth because of the appearance of Jordan blocks associated with the degenerate eigenvalues that merge at that value of $\lambda$, but, as Longhi\cite{op6} has emphasized, this behaviour can be significantly modified depending on the nature of the initial wave packet.

In a previous paper\cite{EMG&HFJ} we approached this problem by explicitly constructing the Bloch wave-functions and the associated Jordan functions corresponding to the degenerate eigenvalues and then using the method of stationary states to construct the $z$-dependence. We found that the explicit linear dependence arising from the Jordan associated functions is indeed cancelled by the combined contributions from the non-degenerate wave-functions and were able to understand how this cancellation came about. In the present paper we approach the problem from a different point of view by revisiting the complementary perturbative calculation of Longhi\cite{op6}.  In Section 2 we briefly recapitulate how the spectrum and eigenfunctions are calculated. Then in Section 3, which forms the main body of the paper, we give an explicit expression for the first-order contribution and carry out the second-order calculation in detail. This enables us to investigate the saturation phenomenon for a variety of different inputs. Finally, in Section 4 we give a brief discussion of our results.

\section{Band structure at threshold}
At the threshold $\lambda=1$, the potential $V$ in Eq.~(\ref{Vsin}) becomes the complex exponential $V=V_0 \exp(2i\pi x/a)$, for which the Schr\"odinger equation is
\bea\label{H1}
-\psi''-V_0\exp(2i\pi x/a)\psi= \beta\psi .
\eea
This is a form of the Bessel equation, as is seen by the substitution $y=y_0 \exp(i \pi x/a)$, where $y_0=(a/\pi)\surd V_0$, giving
\bea\label{Bessel}
y^2\frac{d^2\psi}{dy^2}+y\frac{d\psi}{dy}-(y^2+q^2)\psi=0,
\eea
where $q^2=\beta(a/\pi)^2$. Thus the spectrum is that of a free massive particle, shown in the reduced zone scheme in Fig.~1, and for $q\equiv ka/\pi$ not an integer the solutions $\psi_k(x)=I_q(y)$ and $\psi_{-k}(x)=I_{-q}(y)$ are linearly independent, and have exactly the correct periodicity, $\psi_k(x+a)=e^{ika}\psi_k(x)$, to be the Bloch wave-functions. It is important to note, however, that because the original potential is $PT$-symmetric rather than Hermitian, these functions are not orthogonal in the usual sense, but rather with respect to the $PT$ inner product,
namely
\bea
\int dx \psi_{-k}(x)\psi_{k'}(x)=\delta_{k k'}\int dx \psi_{-k}(x)\psi_k(x),
\eea
\begin{figure}[h]
\resizebox{!}{3in}{\includegraphics{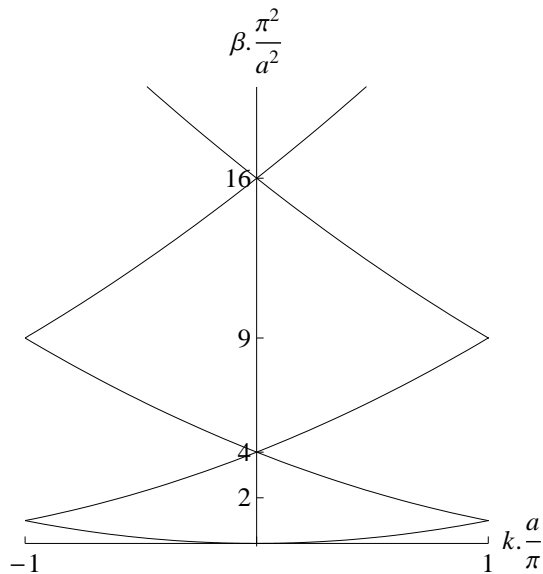}}
\caption{Band structure for $\lambda=1$ in the reduced zone scheme. The Bloch momentum $k$ is plotted in units of $\pi/a$ and the eigengvalue $\beta$ in units of $(a/\pi)^2$.}
\end{figure}
However, for $q=n$, a non-zero integer, $I_n(y)$ and $I_{-n}(y)$ are no longer independent. In that case the Bloch eigenfunctions do not form a complete set, and we must search for other functions, still with the same periodicity, to supplement them. These are the Jordan associated functions, which we denote by $\vf_k(x)\equiv \chi_n(y)$.
They may be defined as derivatives of the eigenfunctions with respect to $\beta$, and satisfy the generalized eigenvalue equation
\bea\label{Jordan}
\left[y^2\frac{d^2}{dy^2}+y\frac{d}{dy}-(y^2+n^2)\right]\chi_n(y)=I_n(y),
\eea
The crucial feature of the Jordan functions is that because of this latter equation they naturally give rise to linear growth in $z$, provided that they are excited:
\bea
e^{-iHz}\vf_r &=& e^{-i\beta_r z} e^{-i(H-\beta_r)z} \vf_r\cr
&=& e^{-i\beta_r z}(\vf_r -i z \psi_r).
\eea
However, as was found numerically in Ref.~\cite{op6}, and explored further in Ref.~\cite{EMG&HFJ}, this natural linear growth may become saturated due to the contributions of neighbouring Bloch functions, which are closely correlated with those of the Jordan functions.

\section{Perturbation Theory}

The analysis of  Ref.~\cite{EMG&HFJ}
 approached the problem from one point of view, in which the interplay between the contributions of the Bloch eigenfunctions and the Jordan associated functions was made explicit. A complementary way of looking at things, which does not separate these two contributions, is to use the perturbative expansion, which instead emphasizes the contributions of the free propagation and the corrections brought about by the potential. The general framework for an expansion of $\psi(x,z)$ in powers of $V_0$, namely $\psi(x,z)=\sum_{r=0}^\infty V_0^r \psi_r(x,z)$, has been given in Ref.~\cite{op6}, along with an approximate form of the first-order term $\psi_1(x,z)$ for the case $q_0=-1$ and $w$ large. In this section we generalize this calculation by obtaining analytic expressions for both the first- and second-order terms for general $q_0$ and $w$. Of course, this can only be used as a guide because there is no guarantee that the expansion converges, nor that the large-$z$ behaviour of the complete amplitude can be extracted from the behaviour of the truncated series.
We will take as our input a Gaussian profile of the form
\bea\label{Gaussian}
\psi(x,0)=f(x)\equiv e^{-(x/w)^2+i k_0 x},
\eea
with offset $k_0$ and width $w$.
The zeroth-order term, $\psi_0(x,z)$, is just the freely-propagating wave-packet
\bea
\psi_0(x,z)=\frac{w}{\surd(w^2+4iz)}e^{ik_0(x-k_0 z)}e^{-(x-2k_0 z)^2/(w^2+4iz)},
\eea
while the first-order term, $\psi_1(x,z)$, is given by
\bea
\psi_1(x,z)=-i\int dk \tilde{f}(k) e^{i(k+k_B)x}\left[-i\int_0^z dy\ e^{-ik^2y}e^{i(k+k_B)^2(y-z)}\right],
\eea
where
\bea
\tilde{f}(k)=\frac{w}{2\sqrt{\pi}}e^{-(k-k_0)^2w^2/4}
\eea
is the Fourier transform of $f(x)$ of Eq.~(\ref{Gaussian}) and $k_B=2\pi/a$ is the width of the first Brillouin zone. We can reverse the order of integration in the expression for $\psi_1$, performing the (Gaussian) $k$ integration first, to obtain
\bea
\psi_1(x,z)&=& -i\frac{w}{\surd{(w^2+4iz)}}e^{\half i k_B x-\quarter i k_B^2 z-\quarter(k_0+\half k_B)^2w^2}\times\\
&&\hspace{3cm}\int_0^z dy\
e^{-\left(2k_By-(x+k_Bz)+\half i w^2(k_0+\half k_B)\right)^2/(w^2+4iz)}.\nonumber
\eea
The $y$ integration is then also a Gaussian integration over a finite range, giving the result
\bea\label{psi1}
\psi_1(x,z)&=& -i\frac{w\sqrt{\pi}}{4k_B}\ e^{\half i k_B x-\quarter i k_B^2 z-\quarter(k_0+\half k_B)^2w^2}\left(\mbox{erf}(\eta_1)+\mbox{erf}(\eta_2)\right),
\eea
where
\bea
\eta_1&=&\frac{k_B z+x-\half i w^2(k_0+\half k_B)}{\surd (w^2+4 i z)}\cr
\eta_2&=&\frac{k_B z-x+\half i w^2(k_0+\half k_B)}{\surd(w^2+4 i z)}\ .
\eea
For the purposes of considering large $w$, it is convenient to rewrite these in the form
\bea\label{arguments}
\eta_1&=&\frac{x-2k_0z}{\surd(w^2+4iz)}\hspace{0.6cm}-\half i(k_0+\half k_B)\surd(w^2+4iz)\cr
\eta_2&=&\frac{2z(k_B+k_0)-x}{\surd(w^2+4iz)}+\half i(k_0+\half k_B)\surd(w^2+4iz)\ .
\eea
The case $k_0=-\half k_B$, i.e. $q_0=-1$, is clearly very special, since in this case the second terms in the contributions to $\eta_1$ and $\eta_2$ vanish, so that we get the simple expressions
$\eta_1=(x+k_B z)/\surd(w^2+4iz)$ and $\eta_2=-(x-k_B z)/\surd(w^2+4iz)$. In that case, as long as $w^2\gg 4 z$ the arguments may be treated as effectively real, and each error function behaves like a sign function of its argument (see Fig.~2(a)), so that the sum of the two behaves like the step function $\theta(k_B z-|x|)$. This is the function $\Phi(x/(k_B z))$ of Ref.~\cite{op6}. In this case the qualitative features of the perturbative calculation are in complete agreement with the spreading of the wave-function in Fig.~3 of that paper, and the saturation\footnote{$\psi_1(x,z)$ grows initially like $z$ for small $z$.} of $\psi_{\rm max}$. However, in this treatment there is of course no mention of whether or not any Jordan functions are excited.

\begin{figure}[h!]
\resizebox{!}{4cm}{\includegraphics{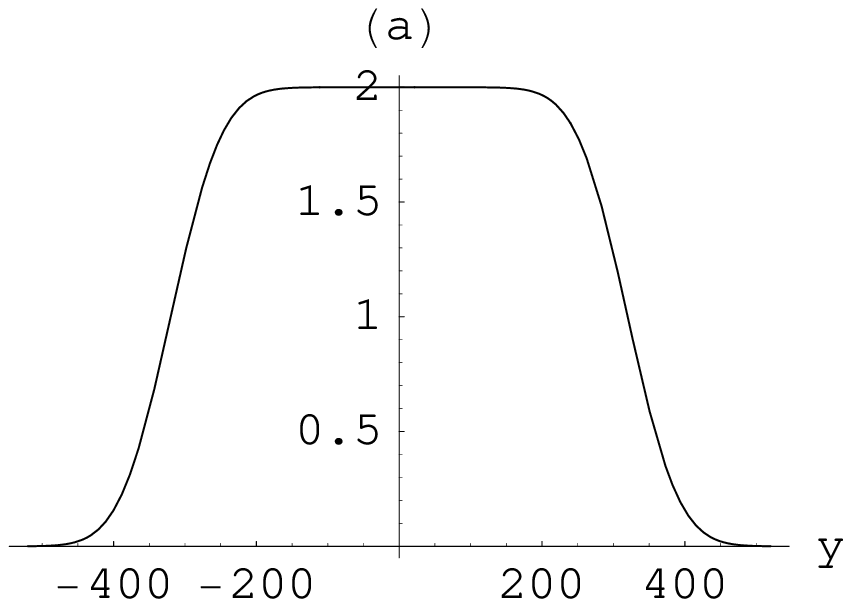}}\hspace{2cm}\resizebox{!}{4cm}{\includegraphics{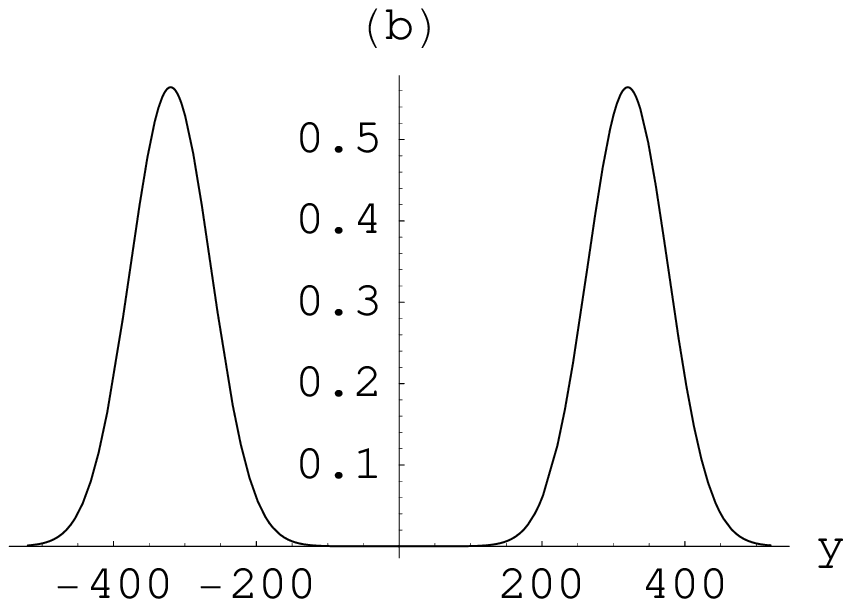}}
\caption{Characteristic behaviour of error functions whose arguments are (a) real or (b) have a large imaginary part. In (a) we plot $\mbox{erf}(4+y/w)+\mbox{erf}(4-y/w)$ with $w=80$. In (b) we plot $w\ e^{-w^2}|\mbox{erf}(4+y/w-i w)+\mbox{erf}(4-y/w+i w)|$ }
\end{figure}
The same expressions in Eqs.~(\ref{psi1}) and (\ref{arguments}) can also be used for the cases $q_0=0$ and $q_0=1$
in the limit of large $w$. In each case the arguments $\eta_1$ and $\eta_2$ now have a large imaginary part. In that situation the modulus of the erf has a narrow peak where the real part vanishes (see Fig.~2(b)). Thus the result consists of two narrow rays, which are centered on $x=2k_0z$ and $x=2(k_B+k_0)z$. Here we have the seeds of the birefringence first observed in Ref.~\cite{op3}. For the case $k_0=q_0=0$, the two rays are centered on $x=0$ and $x=2k_B z$, while
for the case $q_0=1$, or $k_0=\half k_B$, the two rays are centered on
$x=k_B z$ and $x=3k_B z$.

We now go on to second-order perturbation theory to investigate the behaviour of $\psi_2(x,z)$, which is given by\cite{op6}
\bea
\psi_2(x,z)=-\int dk \tilde{f}(k) e^{i(k+2k_B)}\int_0^z d\eta\int_0^{z-\eta} d\xi\  e^{-ik^2z-4ik_b(k+k_B)\eta+ik_B(k_B+2k)\xi}
\eea
Again the $k$ integration is a Gaussian, which leaves finite-range Gaussian integrations over $\xi$ and $\eta$. It is convenient to change the integration variable $\xi$ to $y\equiv 2\eta-\xi$. The integration over $y$ then yields the expression
\bea
\psi_2(x,z)=-\frac{w}{2k_B} \int_0^z d\eta \frac{\sqrt{\pi}}{2}(\mbox{erf}(a)-\mbox{erf}(b))e^{-2ik_B^2\eta}e^{\frac{3}{2}ik_Bx-\quarter ik_B^2t-\quarter w^2\Delta^2},
\eea
where we have written $k_0=-(\half k_B+\Delta)$, and $a$ and $b$ are given by
\bea
b&=&\frac{4 k_B \eta-x-k_B z-\half i w^2\Delta}{\surd{(w^2+4iz)}}\nonumber\\
a&=&\frac{3k_B(2\eta-z)-x-\half iw^2\Delta}{\surd{(w^2+4iz)}}\ .
\eea
The final $\eta$ integrations are then of the form
\bea
I_\eta&\equiv&\int d\eta\ \mbox{erf}(c_1\eta+c_2)e^{c_3\eta}\nonumber\\
&=&\frac{1}{c_3}\left[e^{c_3\eta}\mbox{erf}(c_1\eta+c_2)-e^{c_3(c_3-4c_1c_2)/(4c_1^2)}
\mbox{erf}\left(c_1\eta+c_2-c_3/(2c_1)\right)\right].
\eea
Thus in principle $\psi_2$ is expressible in terms of eight error functions. However, it turns out in practice that only six are involved.

In the case $k_0=-k_B/2$, or $q_0=-1$, when $\Delta=0$, the arguments of the error functions are such as to give a plateau in $|\psi_2|$ between $x=-3 k_B z$ and $x=-k_B z$, i.e. a widening of the beam to the left of $\psi_0$, and a much smaller peak, centered around $x=3k_B z$, that is, a second weak beam to the right.  More importantly, the second-order contribution again shows no sign of the linear growth\footnote{In fact $\psi_2(x,z)$ is proportional to $z^2$ for small $z$.} in $z$ na\"{\i}vely expected from
the excitation of Jordan associated functions.

In the other case,  $q_0=0$, corresponding to $\Delta=-k_B/2$, there are three peaks, centered around $x=0$, $x=-2k_B z$ and $x=4k_B z$, representing a further splitting of the initial beam. Both cases are illustrated in Figure 3.
\begin{figure}[h!]
\begin{center}
\resizebox{!}{4cm}{\includegraphics{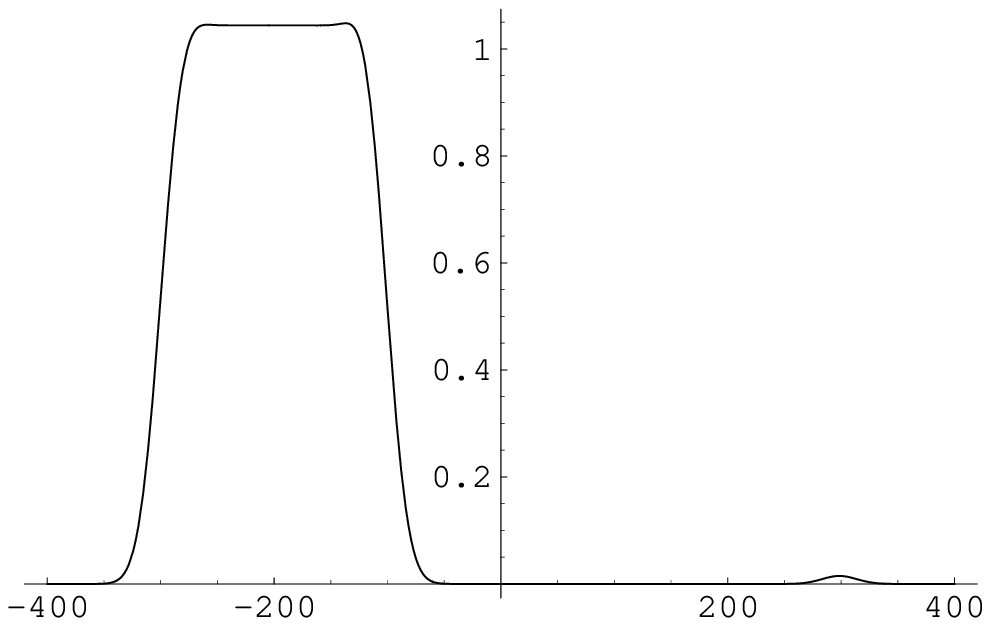}}\hspace{1cm}\resizebox{!}{4cm}{\includegraphics{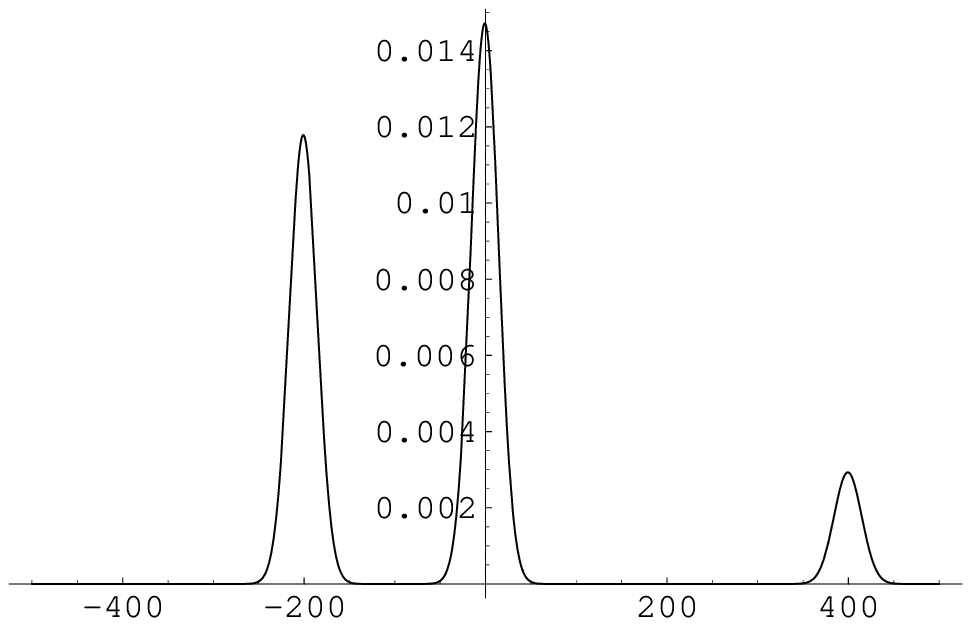}}
\caption{$|\psi_2(x,z)|$ versus $x$ for $z=50$. (a) $q_0=-1$, (b) $q_0=0$.
The parameters are: $a=1$, $V_0=2$ and $w=6\pi$.}
\end{center}
\end{figure}
\section{Discussion}

Of course one needs to treat the results of perturbation theory with caution. In general terms we have no proof that the perturbation series converges, and in particular the asymptotic behaviour in $z$ of a few terms of the series does not necessarily give the correct asymptotic behaviour of the entire sum. Nonetheless this series does appear to give reliable results. For the parameters used by Longhi in Ref.~\cite{op6}, with $V_0=0.2$,  first-order perturbation theory already reproduces the numerical results very well, and the second-order term gives only an extremely small correction.

In all of our calculations we have not allowed $z$ to become too large, restricting it by the condition $z\ll w^2/4$,
in which case saturation is an inbuilt feature of perturbation theory. In fact it was shown in Ref.~\cite{EMG&HFJ} using the method of stationary states that if one goes to much larger values of $z$ the amplitude so calculated  begins to grow again but this ultimate resumption of linear growth is not physical, because it corresponds to the situation where the beam has widened beyond lateral limits of the optical lattice.

It is an elegant feature of the perturbative expansion that the different types of possible behaviour of the beam - spreading or splitting into two or more beams - arise from very simple properties of the error function depending crucially on the offset $k_0$. In the first case the arguments are essentially real, and the error functions behave like sign functions, while in the second case there is a large imaginary part, and the moduli of the error functions behave instead like narrow peaks.

\end{document}